\documentstyle[twoside,fleqn,espcrc2,epsf]{article}

\def\gtwid{\raise.3ex\hbox{$>$\kern-.75em\lower1ex\hbox{$\sim$}}}
\def\ltwid{\raise.3ex\hbox{$<$\kern-.75em\lower1ex\hbox{$\sim$}}}

\def\ie{{\it i.e.},\ }
\def\eg{{\it e.g.},\ }
\def\et{{\it et al.}}

\def\cO{{\cal O}}
\def\cP{{\cal P}}

\def\cZ{{\cal Z}}

\def\nsection#1 #2{\leftline{\rlap{#1}\indent\relax #2}}

\def\prd#1{Phys.\ Rev.\ {\bf D#1}}
\def\plb#1{Phys.\ Lett.\ {\bf #1B}}
\def\npb#1{Nucl.\ Phys.\ {\bf B#1}}

\newcommand{\AmS}{{\protect\the\textfont2
  A\kern-.1667em\lower.5ex\hbox{M}\kern-.125emS}}

\title{Perturbation Theory for Fat-link Fermion Actions}

\author{ C.~Bernard,\hskip-0.03in
\address{{{Department of Physics, Washington
University, St.~Louis, MO 63130, USA}}} 
\thanks{presented by C.~Bernard}
and T.~DeGrand,\hskip-0.03in
\address{Physics Department, University of Colorado, Boulder, CO 80309, USA} %
} 

\begin{document}

\begin{abstract}
We discuss weak coupling perturbation theory for lattice actions in 
which the fermions couple to smeared gauge links.
The normally large integrals that appear in lattice
perturbation theory  are drastically reduced.  Even
without detailed calculation, it is easy to determine to good
accuracy the scale of the logarithms that appear in cases where an
anomalous dimension is present. We describe several 1-loop
examples for  fat-link Wilson and clover fermions.
including the additive mass shift, the relation between the lattice and 
$\overline{MS}$
quark masses, and the axial current renormalization factor ($Z_A$)
for light-light and static-light currents.

\end{abstract}

\maketitle

Smearing or ``fattening'' the gauge links in a fermion action \cite{FATREF}
has several advantages.  Indeed, with sufficient fattening:

\begin{itemize}
\item{} Exceptional configurations are suppressed.

\item{} Additive mass renormalization is small.

\item{} Finite matching constants ($Z$ factors) are very close to 1.

\item{} We expect that tree level $\cO(a)$ improvement  (clover
coefficient
$C_{SW}=1$)
is close to all-orders (in $g$) $\cO(a)$ improvement.
\end{itemize}

The last three features are closely related and can be understood in
perturbation theory.
Furthermore, these actions are already being used in Monte-Carlo computations
(\eg by MILC \cite{MILCfb-lat99}), so it is 
important, as a first step, to have perturbative evaluations
of the renormalization constants.
Non-perturbative evaluations, where possible,  are of course a
very important next step.

  In the present work, we consider the
ordinary Wilson and clover fermion actions
 in which the gauge links are fattened by
APE smearing \cite{APE}. The link after $m+1$ smearings is
related to the link after $m$ smearings by
\begin{eqnarray*}
&V_\mu^{(m+1)}(x) = \cP \Big( (1-c) V_\mu^{(m)}(x)+{}& \\
&{c\over 6}\sum_{\nu\not=\mu}\big[V_\nu^{(m)}(x)
V_\mu^{(m)}(x+\hat\nu)V_\nu ^{\dagger(m)}(x+\hat\mu)&\\
&{}+{}V_\nu^{\dagger(m)}(x-\hat\nu)V_\mu^{(m)}(x-\hat\nu)
V_\nu^{(m)}(x-\hat\nu+\hat\mu)
\big]\Big)\ &
\end{eqnarray*}
where $\cP$ is the projection back into SU(3) and
$V_\mu^0(x) = U_\mu(x)$, the original (``thin'') link.

If one fixes $c$ and performs a fixed number of smearings, $N$, 
as $a\to0$, the action remains local and the continuum limit
will be unaffected.  In practice, MILC has been trying $c=0.45$ 
and $N=10$ for heavy-light decay constants and form factors; we use
these values for all ``fat'' results quoted in this paper. 
(Unless specified otherwise, we take $C_{SW}=1$ in clover computations.)

We are interested in computing 2- and 4-quark operator renormalization/matching
constants at one loop. For such computations, only the linear part of the
relation between fat and thin links is relevant 
($V_\mu^{(m)}(x) \equiv \exp(iaA_\mu^{(m)}(x)$):
\begin{equation}
\label{eq:linA}
A^{(1)}_\mu(x) = \sum_{y,\nu} h_{\mu\nu}(y) A_\nu^{(0)}(x+y)\ .
\end{equation}
Quadratic terms in (\ref{eq:linA}), which would be relevant for tadpole
graphs only, appear as commutators and therefore do not contribute, since
tadpoles are symmetric in the two gluons.

In momentum space, eq.~(\ref{eq:linA}) becomes 
\begin{equation}
\label{eq:momspA}
 A^{(1)}_\mu(q) = \sum_{\nu} \tilde h_{\mu\nu}(q) A_\nu^{(0)}(q)\ ,
\end{equation}
where
\begin{equation}
\label{eq:hmunu}
\tilde h_{\mu\nu}(q) = f(q)(\delta_{\mu\nu} - {\hat q_\mu\hat q_\nu
\over \hat q^2})
+ {\hat q_\mu\hat q_\nu\over \hat q^2}\ ,
\end{equation}
with $\hat q_\mu ={2\over a} \sin({aq_\mu\over2})$ and
$f(q) =  1 - {c\over6}\hat q^2$.

After $N$ smearings, $\tilde h_{\mu\nu}(q)$ becomes
$\tilde h^N_{\mu\nu}(q)$, which is just $\tilde h_{\mu\nu}$
with $f$  replaced by $f^N$. If
$0\le c\le 0.75$, then $|f|\le1$ over the entire
Brillouin zone, and the factor $f^N$ will, for large $N$, strongly
suppress values of $q$ outside of a small ball around $q=0$.

For  small $c$, it is easy to find the effective range of
the fattening:
\begin{equation}
f^N(q) \sim \exp(-{cN\over6}\hat q^2)
 \Rightarrow \langle x^2 \rangle \sim {cN\over3}a^2
\end{equation}
Thus, even an  $N=10$, $c=0.45$ fat link has a mean size of only a couple
of lattice spacings.

The  longitudinal part of $A_\mu$ in 
eq.~(\ref{eq:momspA}) is unaffected by smearing,
because it is a pure gauge,
for which parallel transport is independent of path.

The Feynman rules are easily derived from
 eqs.~(\ref{eq:momspA},\ref{eq:hmunu}).
Each quark-gluon vertex gets a form factor $\tilde
h^{N}_{\mu\nu}(q)$, where $q$ is the gluon momentum.  If all gluon lines 
start and end on fermion lines, then, effectively, the
gluon propagator changes by
$
D_{\mu\nu} \longrightarrow  \tilde h^{N}_{\mu\lambda}D_{\lambda\sigma}
\tilde h^{N}_{\sigma\nu}$.
The form of $\tilde h_{\mu\nu}$ shows
 that Landau gauge is the most natural gauge for fat-link
perturbation theory, since the Landau propagator will kill the
 longitudinal part of  $\tilde h$.

In many cases it is now easy to convert 
ordinary thin-link perturbation theory to the fat-link case.
  In particular, consider 
some pure lattice quantity (\eg the additive mass shift, or
$Z_A$ for two light quarks) computed for
thin links at one loop.  The answer is of the form $\int 
{d^4q \over(2\pi)^4}  I_{latt}(q)$.  If the thin-link calculation was done
in Landau gauge with the loop momentum along the gluon line,
then the fat-link result is obtained simply by replacing
$I_{latt}(q) \to I_{latt}(q) f^{2N}(q)$.

However, if the thin-link calculation was done
in Feynman gauge with the loop momentum along the gluon line,
the fat-link Landau gauge computation will have some new terms, coming
from the $f^{N} {\hat q_\mu\hat q_\nu
\over \hat q^2}$ part of $\tilde h^{N}_{\mu\nu}(q)$.  When
$f=1$, the sum of such terms must vanish by gauge invariance.
If the cancellation occurs at the level of the integrands, 
multiplying by  $f^{2N}$ will not affect it,
and the change from thin to fat links is again accomplished
by $I_{latt}(q) \to I_{latt}(q) f^{2N}(q)$.   
In the cases we have checked (additive mass shift, multiplicative
mass renormalization), the cancellation does indeed
occur at the level of the integrands.  However, we know of no general
proof that this must be the case.  If the cancellation required
integration by parts, for example, the
fat-link computation would contain some new terms not present
in the thin-link case.

We can now perform the lattice integrals numerically.  An important
first example is the simple gluon tadpole ($I_{latt}= {1\over \hat q^2}$).
We find that it has the value 0.1547 in the thin-link case, but only
0.0044 for fat links.
The form factor $f^{2N}$ has completely suppressed the UV part
of the Brillouin zone, which gives the dominant contribution to
the tadpole. 
Thus
 tadpole improvement is
unnecessary for fat links, at least with this much smearing.

Another example is $Z_A^{\rm ll}$, the axial current
renormalization constant for two  massless quarks.
Writing $Z_A^{\rm ll} = 1 + {g^2 C_F\over 16 \pi^2}\zeta^{\rm ll}_A$,
we have 
\begin{equation}
\zeta^{\rm ll}_A = \cases{-15.80,&thin Wilson\cr
                   -13.80,&thin clover\cr
                   -0.24,&fat clover \cr}
\end{equation}
(Here clover currents are local: improved by the factor
$1+m_0 =1 + {1\over2\kappa}-{1\over2\kappa_c}$,
not by derivative terms.) Again, the form factor  has suppressed
most of the integration region, making the integral small. 
With this much fattening, 
all such pure lattice integrals should be highly suppressed.
This explains why the additive mass renormalization is so small,
and why a clover coefficient $C_{SW}=1$ should be very close
to the nonperturbatively improved value --- certainly the tadpole-improved
$C_{SW}$ is extremely close to 1.

The matching of a divergent operator (\ie one with an anomalous dimension)
is slightly more complicated.
In this case, a standard way to express the lattice result in the 
thin case is:
\begin{eqnarray}
&\int {d^4q\over(2\pi)^4} \left[I_{latt}\left(q\right) - I_{cont}\left(q,0\right)\right]& \nonumber\\
&{}+ \int {d^4q\over(2\pi)^4} I_{cont}(q,a\lambda)\ ,
\end{eqnarray}
where $\lambda$ stands for some IR cutoff, such as a quark mass
or momentum, or a gluon mass inserted by hand. 
$I_{cont}(q,a\lambda)$ is a simple continuum-like integrand which
has the same IR behavior as the lattice integrand, but which can be
integrated analytically, giving the explicit $\ln(a\lambda)$  behavior.

In the fat case, the analogous formula is
\begin{eqnarray}
\label{eq:fatdiverge}
&\int {d^4q\over(2\pi)^4} f^{2N}\left[I_{latt}
\left(q\right)
-  I_{cont}\left(q,0\right)\right] &\nonumber\\
&{}+  \int {d^4q\over(2\pi)^4} I_{cont}(q,a\lambda) &\nonumber\\
&{}+\int {d^4q\over(2\pi)^4} I_{cont}\left(q,0\right)
\left(f^{2N}-1\right)\ ,&
\end{eqnarray}
where we have added
and subtracted the integral of $f^{2N} I_{cont}$.  This separates
the result into three terms:  (1) a complicated, regularized
lattice integral, which is however suppressed for
sufficient fattening and is thus numerically small, (2)
the same continuum-like integral that appeared in the thin case,
and (3) a simple integral
involving the form factor,
 whose coefficient is determined
by the anomalous dimension of the operator.  Term (3)
is not particularly small, since $f^{2N}$ does
not multiply the entire integrand.

An example is the matching of the lattice mass to the continuum
${\overline{\rm MS}}$ mass:
\begin{equation}
\cZ_M(a\mu) = 1 + {g^2 C_F\over 16 \pi^2}\left(\zeta_M - 6\log\left(a \mu\right)\right) \ .
\end{equation}
We get
\begin{equation}
\zeta_M = \cases{12.95,&thin Wilson\cr
                   19.31,&thin clover \cr
                   -4.92,&fat clover\ . \cr}
\end{equation}

$Z_A^{\rm sl}$, the axial current renormalization constant in the static-light
case, also has an anomalous dimension.
It does not separate as easily as eq.~(\ref{eq:fatdiverge}).
However, the calculations, following \cite{STATIC}, are straightforward.
We find
\begin{equation}
Z^{\rm sl}_A(aM_B) = 1 + {g^2 C_F\over 16 \pi^2}\left(\zeta_A^{\rm
sl}+ 3\log\left(a M_B\right)\right)
\end{equation}
\begin{equation}
\zeta^{\rm sl}_A = \cases{-22.36,&thin Wilson\cr
                   -16.41,&thin clover\cr
                   0.393,&fat clover\ .\cr
}
\end{equation}
(The clover currents are local.)

The bad news about fat links is that, at least with the amount
of fattening used in the examples above ($N=10$, $c=0.45$),
the lattice integrands are so suppressed at large $q$ that they may become
IR sensitive.  A measure of the effective cutoff of the integrals is given
by $q^*$ \cite{LandM}.  With $C_{SW}=1$,
we find  $q^* a =0.71$ for $Z_A^{\rm ll}$; while
$q^* a = 1.05$ for 
$\Sigma_0$ (which gives the shift in $\kappa_c$).

More
direct evidence that perturbation theory for $\Sigma_0$ is breaking down
comes from a comparison with numerical results from the MILC collaboration.
For $\beta\!=\!5.6$, $m\!=\!.01$, $C_{SW}\!=\!1.0$,
$N\!=\!10$, $c\!=\!0.45$, the numerical result is $\kappa_c=0.1256(1)$;
whereas 1-loop perturbation theory, with $q^*a \!=\! 1.05$, gives
$\kappa_c=0.1251$.  Thus perturbation theory fails by a factor of six
for $\kappa_c-1/8$. There are however sensitive cancellations at 1-loop
making $\Sigma_0$ anomalously small.

For the same data set, $Z^{\rm ll}_A=0.99$ in perturbation theory.
We guess this will be better behaved than $\Sigma_0$, because there
are no delicate cancellations here.
But even if
$Z^{\rm ll}_A-1$ is wrong by another
factor of 6,  the associated error in the final
answer is only 5\%, since $Z^{\rm ll}$ is so close to 1.

However, in cases with anomalous dimensions (\eg $f_B$), 
the perturbative corrections are
not necessarily small, and a failure of perturbation theory
could lead to large errors.
Some non-perturbative calculations are needed to clarify this situation.

In retrospect, somewhat less fattening in the MILC running
would probably have been preferable.
For example, for the clover $Z_A^{\rm ll}$,  
$N=7$, $c=0.25$ gives $q^* a =1.34$ instead of
$0.71$, and may still
suppress exceptional configurations enough.

We thank C.\ DeTar and M.\ Stephenson for sharing their
numerical results with us, and all of our MILC colleagues
for discussions.
This work was supported in part by the DOE.


\begin{thebibliography}{9}
\vspace{-4pt}

\bibitem{FATREF}
The approaches to fattening which most influenced us
are
  T.\ DeGrand \et, \npb{547} (1999) 259; T.\ Blum \et, \prd{55} (1997) R1133.

\bibitem{MILCfb-lat99}  C.\ Bernard \et, 
these proceedings.

\bibitem{APE}
M.\ Albanese \et,
\plb{192} (1987) 163.

\bibitem{STATIC}
E.\ Eichten and B.\ Hill,
\plb{240} (1990) 193; 
A.\ Borrelli and C.\ Pittori,
\npb{385} (1992) 502;
O.\ Hernandez and B.\ Hill, \plb{289} (1992) 417.

\bibitem{LandM}
G.P.\ Lepage and P.\ Mackenzie, \prd{48} (1993) 2250.
\end{thebibliography}
\end{document}